\documentclass[11pt]{llncs}
\usepackage{graphicx}
\usepackage{cite}
\usepackage{url}
\usepackage{times}
\usepackage{mathptm}

\DeclareSymbolFont{AMSb}{U}{msb}{m}{n}
\DeclareSymbolFontAlphabet{\Bbb}{AMSb}

\def\nr{{\mathop{\rm nr}}}
\def\usp{{\mathop{\rm usp}}}
\def\ssg{{\mathop{\rm ssg}}}
\def\lnot{{\mathop{\sim}}}

\let\oldendproof\endproof
\def\endproof{\qed\oldendproof}

\begin{document}

\title{Nonrepetitive Paths and Cycles in Graphs\\
with Application to Sudoku} 

\author{David Eppstein}

\institute{Computer Science Department\\
Donald Bren School of Information \& Computer Sciences\\
University of California, Irvine\\
\email{eppstein@uci.edu}}

\maketitle   

\begin{abstract}
We provide a simple linear time transformation from a directed or undirected graph with labeled edges to an unlabeled digraph, such that paths in the input graph in which no two consecutive edges have the same label correspond to paths in the transformed graph and vice versa.  Using this transformation, we provide efficient algorithms for finding paths and cycles with no two consecutive equal labels.  We also consider related problems where the paths and cycles are required to be simple; we find efficient algorithms for the undirected case of these problems but show the directed case to be NP-complete. We apply our path and cycle finding algorithms in a program for generating and solving Sudoku puzzles, and show experimentally that they lead to effective puzzle-solving rules that may also be of interest to human Sudoku puzzle solvers.
\end{abstract}

\section{Introduction}

In an edge-labeled directed or undirected graph, we say that a path or cycle (not necessarily simple) is {\em nonrepetitive} if no two consecutive edges share the same label.  We are interested in this paper in finding such nonrepetitive paths and cycles.  We provide the following new results:

\begin{itemize}
\item We show how to transform any input graph $G$ with $n$ vertices and $m$ labeled edges into a new directed graph $\nr(G)$ with $O(m)$ edges and vertices, such that any nonrepetitive path in $G$ corresponds to a path in $\nr(G)$ and vice versa.  The transformation can be performed in linear time, if we assume that the edges at each vertex of the input are grouped according to their labels
(in practice this grouping can achieved efficiently using a hash table if it does not exist already).  By using strong connectivity analysis in $\nr(G)$, we show how to identify, in linear time, all edges of $G$ that belong to nonrepetitive cycles.  By using depth first search in $\nr(G)$ we identify, given a vertex $v$ and label $L$, all edges of $G$ that belong to a nonrepetitive path starting from $v$ with initial edge label $L$, again in linear time.

\item We provide efficient algorithms for finding nonrepetitive simple paths in undirected graphs.  In linear time, we can test whether there exists such a path between any two vertices of the input graph.
In time $O(m^2)$, we can list all edges of the graph that belong to nonrepetitive simple cycles.

\item We show that, for edge-labeled directed graphs, it is NP-complete to test for the existence of a nonrepetitive simple path between two vertices, or to test whether a given edge belongs to a nonrepetitive simple cycle.

\item We consider a rule-based approach to solving Sudoku puzzles~\cite{Wik-Sudoku} that attempts to model human reasoning more accurately than would a brute-force backtracking search for the solution.  We define several polynomial time puzzle-solving rules for Sudoku based on our algorithms for nonrepetitivity analysis in graphs, and validate experimentally their effectiveness at solving randomly generated Sudoku puzzles.
\end{itemize}

Readers interested primarily in Sudoku may skip directly to section~\ref{sec:sudoku}. In the remainder of this section we discuss prior research related to our concept of nonrepetitivity, and in Section~\ref{sec:nonrep} we describe in detail our algorithmic results on nonrepetitive paths and cycles.

\subsection{Prior Work}

We are not aware of previous work on our version of nonrepetitivity, but there has been some research on related concepts of repeated labels in paths and cycles.  In previous work on certain number-theoretic algorithms~\cite{Epp-MER-95} we noted that the problem of finding a path between two vertices that uses each label at most once (regardless of consecutivity) is NP-complete; this can be proven by a simple reduction from 3SAT.  In that paper, we found such paths by resorting to a heuristic search that lists short paths until one without repetitions is found; alternatively, if the number of possible labels is a small number $k$, it is straightforward to find such paths by dynamic programming in the fixed-parameter-tractible time bound $O(2^k m)$.  The idea of avoiding repeated labels also occurs in the color coding technique of Alon, Yuster, and Zwick~\cite{AloYusZwi-JACM-95} for finding long paths and cycles in unlabeled graphs.

Another concept closely related to nonrepetitivity is that of alternating paths in graph matchings; indeed, an alternating path is just a nonrepetitive simple path in a  graph in which edges have been labeled as belonging to or not belonging to a given matching, so Edmonds' blossom contraction algorithm~\cite{Edm-CJM-65} can be seen as a special case of searching for nonrepetitive paths in undirected graphs.   Our results on simple paths show that alternating path search algorithms can be extended more generally to the case of undirected nonrepetitive simple paths, but not the case of directed nonrepetitive simple paths.

\section{Nonrepetitive Paths and Cycles}
\label{sec:nonrep}

In this section we describe our results on algorithms for finding nonrepetitive paths and cycles in arbitrary graphs.  We begin with algorithms for finding paths and cycles that need not be simple.

\subsection{Paths and Cycles Without Requiring Simplicity}

Our algorithms for finding paths and cycles that need not be simple operate by replacing each vertex by a {\em gadget}, a subgraph that allows paths to connect through the vertex only when the path edges that enter and exit the vertex have different labels. The obvious way to achieve this is to use a gadget consisting of a pair of vertices per label, one for incoming edges and one for outgoing edges with that label, and to connect each incoming vertex to each differently-labeled outgoing vertex. However, this would result in nonlinear size for the replacement graph when vertices can have many labels incident to them.  It is possible, however, to replace vertices by somewhat more complex gadgets that have size proportional to the number of labels at each vertex; this technique leads to a linear overall size bound for our replacement graph.

\begin{figure}[t]
\begin{center}
\includegraphics[width=2.5in]{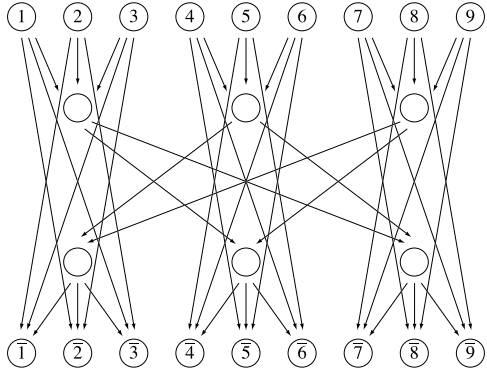}
\end{center}
\caption{Gadget for transforming nonrepetitive paths into unlabeled paths.  Each of the nine labeled vertices at the top can reach eight of the nine vertices at the bottom, excluding the one with the matching label.}
\label{fig:gadget}
\end{figure}

\begin{lemma}
\label{lem:gadget}
For any $k$, there is a directed graph $D$ with $O(k)$ vertices and edges, one vertex labeled $x$ for each integer $1\le x\le k$, and one vertex labeled $\bar x$ for each integer $1\le x\le k$, such that $x$ can reach $\bar y$ in $D$ if and only if $x\ne y$.
\end{lemma}

\begin{proof}
For $D=2$, the result is clear: simply use four vertices, and connect $1$ to $\bar 2$ and $2$ to $\bar 1$.
For larger $D$, we construct by induction a graph $D'$ with labels in the range $1\le x\le\lceil k/2\rceil$.
For each $1\le x\le k$, let $y=x+1$ if $x$ is odd, $y=x-1$ if $x$ is even, and $z=\lceil x/2\rceil$.  We connect
$x$ to $\bar y$, $y$ to $\bar x$, both $x$ and $y$ to the vertex labeled $z$ in $D'$,
and the vertex labeled $\bar z$ in $D'$ to $\bar x$ and $\bar y$.  It is easily verified by induction that the resulting graph satisfies the requirements of the lemma, and has $O(k)$ vertices and edges.
\end{proof}

The implementation of these algorithms that we use in our Sudoku program uses a similar but slightly more complicated construction that, when possible, groups labels into triples instead of pairs; an example of this triple grouping gadget for labels consisting of the nine digits of a Sudoku puzzle is shown in Figure~\ref{fig:gadget}.

We are now ready to define $\nr(G)$.  Suppose we are given a graph $G$ with labeled edges.
For each vertex $v$ in $G$, use Lemma~\ref{lem:gadget} to create a gadget in $\nr(G)$ for $v$,
with the number $k$ of incoming and outgoing labels in the gadget equal to the number of distinct labels on edges incident to $v$ in $G$.  For each edge $(v,w)$ labeled $x$ in $G$, create an edge in $\nr(G)$ from the vertex labeled $\bar x$ in $v$'s gadget to the vertex labeled $x$ in $w$'s gadget.  This construction results in a graph with $O(m)$ vertices and edges; it can be performed in time $O(m)$ if the input graph $G$ has its edges grouped by label at each vertex.  If this grouping does not already exist, it can be found efficiently using hashing.  The following result follows immediately from the properties of the replacement gadgets used to construct $\nr(G)$:

\begin{lemma}
If $\pi$ is a path in $\nr(G)$, the sequence of gadgets through which $\pi$ passes corresponds to a nonrepetitive path in $G$.  Conversely, if $\pi$ is a nonrepetitive path in $G$, there exists a path in $\nr(G)$ that passes through a sequence of gadgets corresponding to the vertices of $\pi$.
\end{lemma}

\begin{theorem}
\label{thm:all-nr}
Given a labeled graph, in $O(m)$ time we can identify all edges that belong to nonrepetitive cycles of~$G$.
\end{theorem}

\begin{proof}
We construct $\nr(G)$ and compute its strongly connected components.  Edge $(v,w)$ with label $x$ belongs to a nonrepetitive cycle in $G$ if and only if the vertex labeled $\bar x$ in $v$'s gadget in $\nr(G)$ and the vertex labeled $x$ in $w$'s gadget in $\nr(G)$ belong to the same strongly connected component, which can be tested in constant time per edge if we label each vertex in $\nr(G)$ by the identity of the component containing it.
\end{proof}

\begin{theorem}
\label{thm:reach}
Given a labeled graph, a vertex $v$, and a label $x$, in $O(m)$ time we can identify all edges that belong to nonrepetitive paths of $G$ that have $v$ as their initial vertex and $x$ as their initial edge label.
\end{theorem}

\begin{proof}
We construct $\nr(G)$ and use depth first search to identify all vertices reachable from the vertex labeled $\bar x$ in $v$'s gadget in $\nr(G)$.  It is possible for a nonrepetitive path in $G$ to reach edge $(u,w)$ with label $y$, starting from $v$ and with initial label $x$, if and only if the vertex labeled $\bar y$ in $u$'s gadget is one of the vertices reached by the search.
\end{proof}

Due to the path-preserving nature of our transformation, we can also use it as part of efficient nonrepetitive versions of other path algorithms: for instance, shortest nonrepetitive paths, $k$-shortest paths, shortest cycles, etc.  The hierarchical nature of our vertex gadgets allows them to be updated by edge insertions or deletions that may change the label sets at vertices, in logarithmic time per update.  We omit the details.

By applying our transformation to an undirected graph $G$, in which we label each edge by its own identity, we obtain a directed graph $\nr(G)$ such that paths in $\nr(G)$ correspond to paths in $G$ that do not traverse the same edge consecutively in opposite directions.  This application of our nonrepetitivity analysis may be of use for path problems in undirected graphs with negative weights but no nontrivial negative cycles.

\subsection{Undirected Simple Paths and Cycles}

\begin{figure}[t]
\begin{center}
\includegraphics[width=4in]{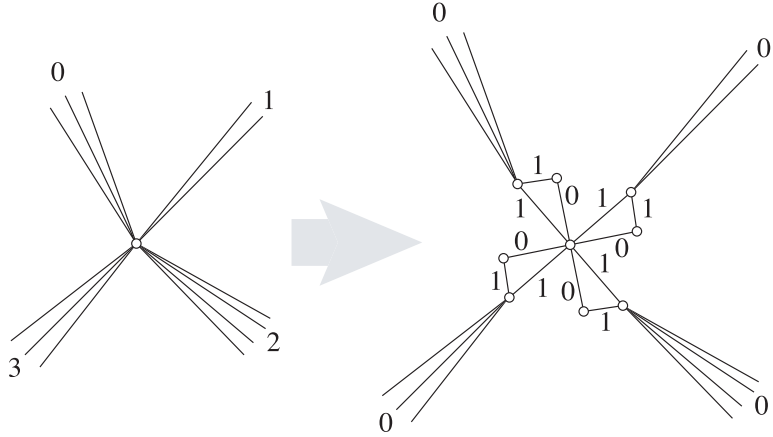}
\end{center}
\caption{Gadget for conversion of an undirected nonrepetitive simple path reachability problem with multiple edge labels into one with only two labels, via Lemma~\ref{lem:binary}.}
\label{fig:binary}
\end{figure}

The requirement that paths be simple (that is, each vertex occurs at most once) is unimportant for our Sudoku applications, but it seems to make the problem of finding non-repetitive paths more complex, even in undirected graphs.  However, such paths can still be found efficiently.  To begin with, we describe a gadget similar to that of Lemma~\ref{lem:gadget} that allows us to assume that our graph's edges only have two labels (which by convention we can assume are the numbers 0 and~1).

\begin{lemma}
\label{lem:binary}
For any labeled undirected graph $G$, we can construct in linear time a transformed graph $\usp(G)$,
such that the edges of $\usp(G)$ have only two labels, 0 and~1, and such that simple nonrepetitive paths in $G$ correspond to simple nonrepetitive paths in $\usp(G)$ and vice versa.
\end{lemma}

\begin{proof}
We replace each vertex of $G$, connected to edges with $k$ labels, by a gadget with $2k+1$ vertices in $\usp(G)$:
$(v,i,0)$, $(v,i,1)$, and $v$.  For each edge $(v,w)$ labeled $i$ in $G$, we connect $(v,i,0)$ and $(w,i,0)$ in $\usp(G)$ by an edge labeled 0.  In addition, we connect $v$ to each $(v,i,0)$ with label $1$, and to each $(v,i,1)$ with label 0, and we connect $(v,i,0)$ to $(v,i,1)$ with label 1.

Then, for any nonrepetitive path $\pi$ in $G$, we can find a nonrepetitive path in $\usp(G)$ as follows:
for each edge in $\pi$, include the corresponding edge in $\usp(G)$, and for each consecutive pair of edges through $v$ labeled $i$ and $j$, include the path through vertices $(v,i,0)$, $v$, $(v,j,1)$, and $(v,j,0)$.  This clearly forms a nonrepetitive simple path in $\usp(G)$.  Conversely, for any nonrepetitive simple path in $\usp(G)$, we can form a path in $G$ by keeping only the edges in $G$ corresponding to edges of the form $(v,i,0)$ to $(w,i,0)$ in $\usp(G)$. Whenever such a path passes through a vertex gadget in $\usp(G)$, it must pass through the central vertex $v$, and simplicity prevents it from leaving $v$ through the same vertex $(v,i,0)$ from which it entered.
\end{proof}

In Figure~\ref{fig:binary} we depict the gadgets used to replace each vertex in the proof of Lemma~\ref{lem:binary}.

Goldberg and Karzanov~\cite{GolKar-SODA-94} define a {\em skew-symmetric graph} to be a directed graph, together with an involution $\sigma$ on its vertices that reverses the orientation of each edge.
They define the $r$-reachability problem to be one of finding a path in a skew-symmetric graph from a given vertex $s$ to $\sigma(s)$ such that, for each vertex $v$ of the graph, at most one of $v$ and $\sigma(v)$ is used in the path.  Such a path may be assumed to be simple, for any loops can be removed without violating the skew-symmetric condition.  As we now show, nonrepetitive simple paths in undirected graphs with binary edge labels can be transformed into a skew-symmetric reachability problem.

\begin{lemma}
\label{lem:skewsym}
Let $G$ be an undirected graph in which the edges are labeled by 0 and 1, and let $p$ and $q$ be two chosen vertices in $G$.  Then in linear time we can construct four skew-symmetric graphs, and nodes $s$ in these graphs, such that $s$ and $\sigma(s)$ are $r$-reachable in one of the two graphs if and only if there exists a nonrepetitive simple path in $G$ from $p$ to $q$.
\end{lemma}

\begin{proof}
For each vertex $v$ in $G$ we construct two vertices $(v,0)$ and $(v,1)$ in $\ssg(G)$, with
$\sigma((v,i))=(v,1-i)$.  For each edge $(v,w)$ labeled 0 in $G$ we connect $(v,1)$ to $(w,0)$ and $(w,1)$ to $(v,0)$ in $\ssg(G)$.  For each edge $(v,w)$ labeled 0 in $G$ we connect $(v,0)$ to $(w,1)$ and $(w,0)$ to $(v,1)$ in $\ssg(G)$.  Additionally we create new vertices $s$ and $\sigma(s)$ in $\ssg(G)$.
To test whether there exists a simple nonrepetitive path from $p$ to $q$ starting and ending on the label~0, we connect $s$ to $(p,1)$ and $(q,1)$, and connect $(p,0)$ and $(q,0)$ to $\sigma(s)$; such a path exists if and only if $s$ and $\sigma(s)$ are $r$-reachable.  The same construction with a slightly different choice of connections from $s$ and $\sigma(s)$ allows testing for paths starting and ending with the other three combinations of labels.
\end{proof}

\begin{figure}[t]
\begin{center}
\includegraphics[width=4in]{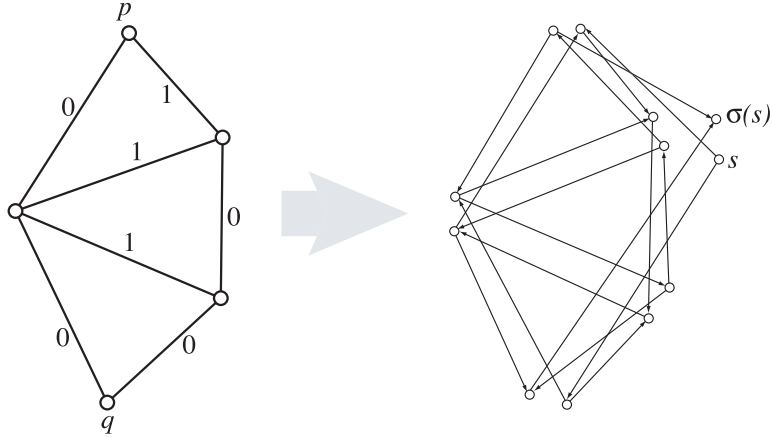}
\end{center}
\caption{Conversion of an undirected binary nonrepetitive simple path reachability problem into a skew-symmetric $r$-reachability problem, via Lemma~\ref{lem:skewsym}.}
\label{fig:skewsym}
\end{figure}

One of the four skew-symmetric graphs constructed in the proof of Lemma~\ref{lem:skewsym} is shown in Figure~\ref{fig:skewsym}.

\begin{theorem}
Given any undirected edge-labeled graph, and vertices $p$ and $q$, we can test in linear time whether the graph contains a nonrepetitive simple path from $p$ to $q$.
\end{theorem}

\begin{proof}
We use Lemma~\ref{lem:binary} to reduce the problem to one with only two labels, use Lemma~\ref{lem:skewsym} to transform it into one of skew-symmetric $r$-reachability, and apply Goldberg and Karzanov's linear time skew-symmetric $r$-reachability algorithm.
\end{proof}

To test whether an edge $(u,v)$ belongs to a nonrepetitive simple cycle, we may remove the edge and its incompatible neighboring edges from the graph and apply this reachability test.  In this way, we can find all edges belonging to nonrepetitive simple cycles, in time $O(m^2)$.

A faster time bound is possible to test for the existence of a single nonrepetitive simple cycle, following closely related algorithms of Gabow et al.~\cite{GabKapTar-STOC-99} and Cook~\cite{Coo-NCCA-03}.  Any vertex in a graph with binary edge labels that is incident only to edges with a single label can not be part of such a cycle, and can be removed; also, any bridge in the graph can not be part of a simple cycle (repetitive or no) and can be removed.  If repeatedly removing single-label vertices and bridges leaves a nontrivial remaining graph, it must contain a nonrepetitive simple cycle; we omit the details.  As Gabow et al.~\cite{GabKapTar-STOC-99} show, dynamic graph algorithms can be used to implement this removal process in time $O(m\log^{O(1)} n)$.  Unfortunately, it is possible that, even after the removal process terminates, there exist edges that do not belong to any nonrepetitive simple cycle.  Perhaps there is some way of combining this idea with cycle contraction to find all cyclic edges more efficiently.

\subsection{Directed Simple Paths and Cycles}

\begin{figure}[t]
\begin{center}
\includegraphics[width=3in]{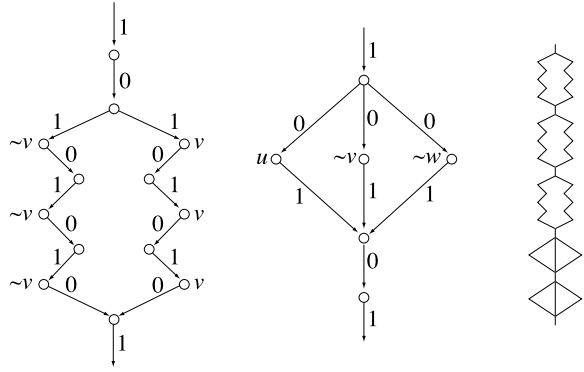}
\end{center}
\caption{Gadgets for proof of NP-completeness of directed nonrepetitive simple path problems. Left: variable gadget; center: clause gadget; right: connecting the gadgets.}
\label{fig:dirnpc}
\end{figure}

Lemma~\ref{lem:binary} extends easily to directed graphs.  Unfortunately, the rest of the algorithms in the previous section do not.

\begin{theorem}
\label{thm:dirnpc}
It is NP-complete, given a directed graph with binary edge labels and two specified vertices $p$ and $q$, to determine whether the graph contains a nonrepetitive simple path from $p$ to $q$, or whether an edge from $q$ to $p$ belongs to a simple nonrepetitive cycle.
\end{theorem}

\begin{proof}
We provide a standard NP-completeness reduction from 3SAT, satisfiability of Boolean formulas in conjunctive normal form with at most three variables per clause.  That is, given such a formula, we construct a labeled digraph such that there exists a nonrepetitive simple path if and only if the formula is satisfiable.

The graph we construct has a subgraph for each variable~$v$, a {\em variable gadget}, shown in Figure~\ref{fig:dirnpc}, left.  This subgraph has incoming and outgoing edges labeled~1, and can connect these two edges along either of two paths.  The vertices with incoming 1-edges and outgoing 0-edges along these two paths are labeled $v$ along one path and labeled $\lnot v$ along the other path.
We also include in our construction, for each clause of the form e.g. $(u\cup\lnot v\cup\lnot w)$, a {\em clause gadget}, shown in Figure~\ref{fig:dirnpc}, center.  This gadget again has incoming and outgoing 1-edges, which can be connected through any of three paths, each of which goes through one of the vertices with the appropriate label from a variable gadget.  Each such labeled vertex should occur in at most one clause gadget, so the paths in the variable gadgets should be made long enough to create as many labeled vertices as will be needed.

The overall structure of the constructed graph connects all these gadgets in an arbitrary sequence by their incoming and outgoing 1-edges.  A simple path from the start to the end vertex in this graph thus chooses one path for each gadget; we cannot form paths that jump out of sequence from variable gadgets to clause gadgets or vice versa because the nonrepetitivity constraint disallows such paths.
Thus, a simple nonrepetitive path exists if and only if we can choose an unused variable label per clause gadget, which happens if and only if we can choose the paths through the variable gadgets in such a way that each clause gadget has an unused variable label, which happens if and only if we can assign truth values to the variables in such a way that each clause has a true variable.

To modify the proof to handle the question of whether an edge belongs to a simple nonrepetitive cycle, simply add an edge from $q$, the end of the path constructed above, back to $p$, and test that edge.
\end{proof}

\section{Sudoku}
\label{sec:sudoku}

Sudoku~\cite{Wik-Sudoku} is a popular puzzle, printed daily in newspapers in Japan, the United Kingdom, and the USA, in which the aim is to fill a $9\times 9$ matrix of cells with digits from $1$ through~$9$.  Each puzzle consists of a grid in which some digits have already been filled in, and the goal is to fill in the remaining cells so that each digit appears once in each row, once in each column, and once in each of nine $3\times 3$ squares into which the grid has been subdivided.  Several (difficult) examples of such puzzles are shown in the figures of this section.

A proper Sudoku puzzle must have a unique solution, and it should be possible to reach that solution by a sequence of logical deductions without trial and error.  Although Sudoku, when generalized to $B^2\times B^2$ grids to be filled in by numbers from $1$ to $B^2$, is NP-complete~\cite{YatSet-IPSJ-02}, it is not difficult for a computer program to solve most Sudoku puzzles quickly by a brute force backtracking search.  However, it is of interest to instead attempt to mimic human Sudoku solvers, and derive rules that solve Sudoku puzzles without backtracking, first for the standard AI reason that such attempts can teach us much about the power of human and machine reasoning, and second because human-like problem solving capabilities allow us to automatically estimate the difficulty of Sudoku puzzles  for human solvers by examining the rules necessary to solve each puzzle.  In particular, the more accurately we can model the power of human-like reasoning by a rule-based Sudoku solver, the more accurately we will be able to automatically distinguish ``proper'' puzzles that allow human step-by-step solution from ``improper'' puzzles that seemingly require trial and error for their solution.

To model the requirement of step-by-step deductive solution, we first require that all rules used by our solver must be implementable as algorithms that, when generalized to $B^2\times B^2$ sized puzzles, take time polynomial in~$B$.  Since in actual Sudoku puzzles $B$ is a small number, three, we can tolerate polynomials of $B$ with moderately large exponents in our runtimes.  However, this polynomial time restriction would not by itself prevent us from including rules that perform a limited amount of backtracking, such as experimentally filling one cell and then applying other rules in sequence to determine whether that choice leads to an inconsistency.  To avoid such limited backtracking rules, we also demand (although this is less mathematically well defined) that our rules perform a single action, typically searching for some pattern that will allow us to resolve whether or not some digit can be placed in some cell, rather than merely determining consequences of tentative decisions.

\subsection{Local Rules}
\label{sec:local}

In the Sudoku solving program we implemented, many of the puzzle solving rules are {\em local}; that is, they examine only one or two digits, rows, columns, or squares of the puzzle at a time, rather than considering the puzzle as a whole.  In order to place our new nonlocal rules in the context of our solver, and make sense of our experimental results, we briefly describe the other local rules we have implemented:

\begin{itemize}
\item If a digit $x$ has only one remaining cell that it can be placed in, within some row, column, or square, then we place it in that cell. Any potential positions of $x$ incompatible with that cell (because
they lie in the same row, column, or square) are removed from future consideration.

\item If a cell has only one digit $x$ that can be placed in it, we place $x$ in that cell.  Incompatible positions for $x$ are removed from future consideration.

\item If some three cells, formed by intersecting a row or column with a square, have three digits whose only remaining positions within that row, column, or square are among those three cells,
we prevent all other digits from being placed there.  We also remove positions for those three forced digits outside the triple but within the row, column, or square containing it.

\item If the cells of a square that can contain a digit $x$ all lie in a single row or column, we eliminate positions for $x$ that are outside the square but inside that row or column.  Similarly, if the cells that can contain $x$ within a row or column all lie in a single square, we eliminate positions that are inside that
square but outside the row or column.

\item If two digits $x$ and $y$ each share the same two cells as the only locations they may be placed within some row, column, or square, then all other digits must avoid those two cells.

\item If the placement of digit $x$ in cell $y$ can not be extended to a placement of nine copies of $x$ covering each row and column of the grid exactly once, we eliminate cell $y$ from consideration as
a placement for~$x$.

\item If the placement of a digit $x$ in cell $y$ within a single row, column, or square can not be extended to a complete solution of that row, column, or square, then we eliminate that placement from consideration.
\end{itemize}

Among these rules, the last two are the most interesting from an algorithmic point of view.  In both cases, they can be performed using an algorithm based on perfect matching in bipartite graphs.  For instance, a placement of nine copies of the digit $x$ can be viewed as a perfect matching in a graph in which the vertices are the rows and columns of the grid, and in which two vertices are connected by an edge when $x$ can be placed in the cell shared by the row and column corresponding to those vertices.  By finding a single perfect matching, orienting the edges of the graph from rows to columns on matched edges and from columns to rows on unmatched edges, and selecting the unmatched edges whose endpoints belong to different strongly connected components of this oriented graph, we can find the positions where a placement of one copy of $x$ can not be extended to a complete placement of all nine copies.  Similarly, solving a subproblem consisting a single row, column, or square can be represented as finding a perfect matching in a graph the vertices of which correspond to the nine digits and the nine cells of the row, column, or square, and a similar strong connectivity analysis finds the placements that can not be part of such a solution.  In generalized $B^2\times B^2$ Sudoku grids, both of the final two rules involve computing perfect matchings and strongly connected components in $O(B^2)$ graphs, each of which has $2B^2$ vertices, and can therefore be performed in total time~$O(B^7)$.

The connection with matching also allows us to reformulate the final two rules via Hall's theorem, in a way that appears superficially less general but closer to our requirement that each rule search for a pattern that a human could be expected to find:

\begin{itemize}
\item If there exist sets $S$ of rows and $T$ of equally many columns, such that the only cells among rows $S$ where digit $x$ can be placed also belong to the columns in $T$, then $x$ can not be placed in a cell that lies in a column of $T$ but does not lie in a row of $S$.
\item If for some row, column, or square $g$ there exists a set $S$ of digits, and a set $T$ of equally many cells of~$g$, such that the digits of $S$ can only be placed in~$g$ within the cells of $T$,  then only digits of $S$ can be placed in the cells of $T$.
\end{itemize}

These two rules could also be restated, reversing the roles of rows and columns in the first rule, and of digits and cells in the second, but that would not add any generality or solving power to them.  This reinterpretation makes clear that several of our earlier rules are subsumed by our two matching rules; however, it is useful to include the simpler rules in our solver both for computing speed and to allow us to classify puzzles as more easily solvable.

\subsection{Nonrepetitive Cycle Rule}

\begin{figure}[t]
\begin{center}
\includegraphics[width=2.25in]{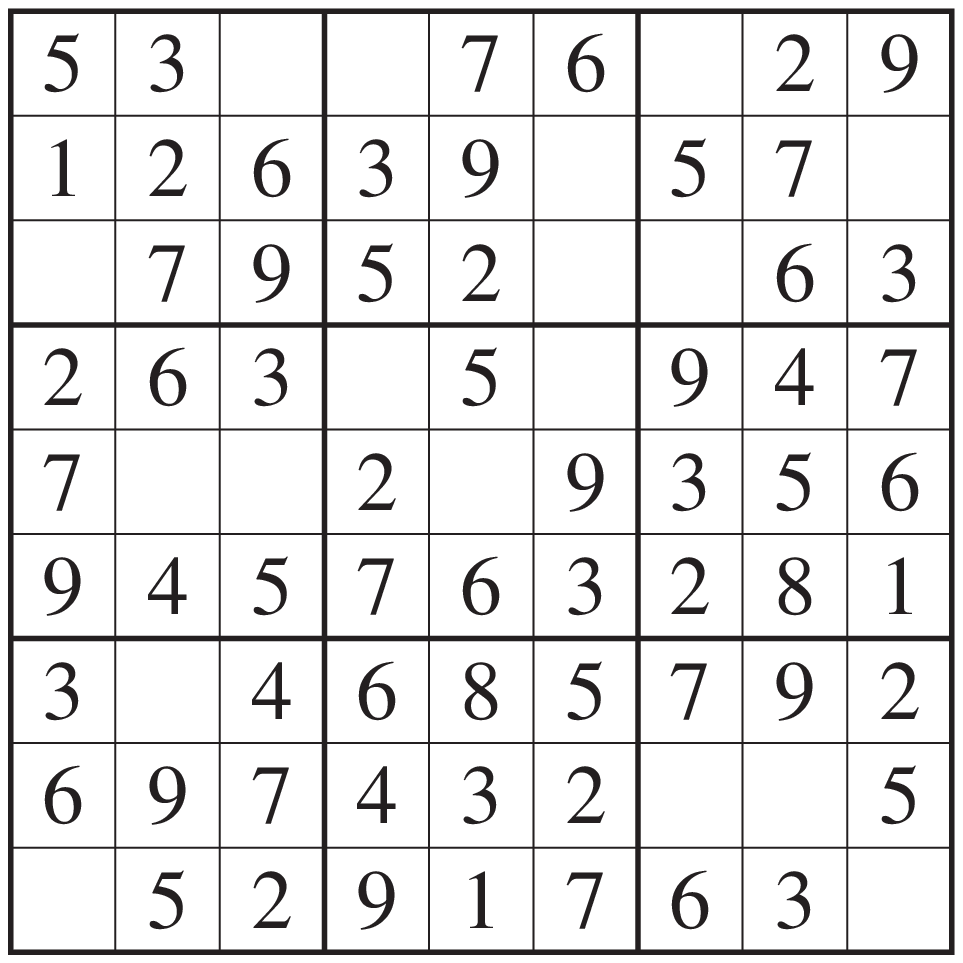}
\hspace{0.25in}
\includegraphics[width=2.25in]{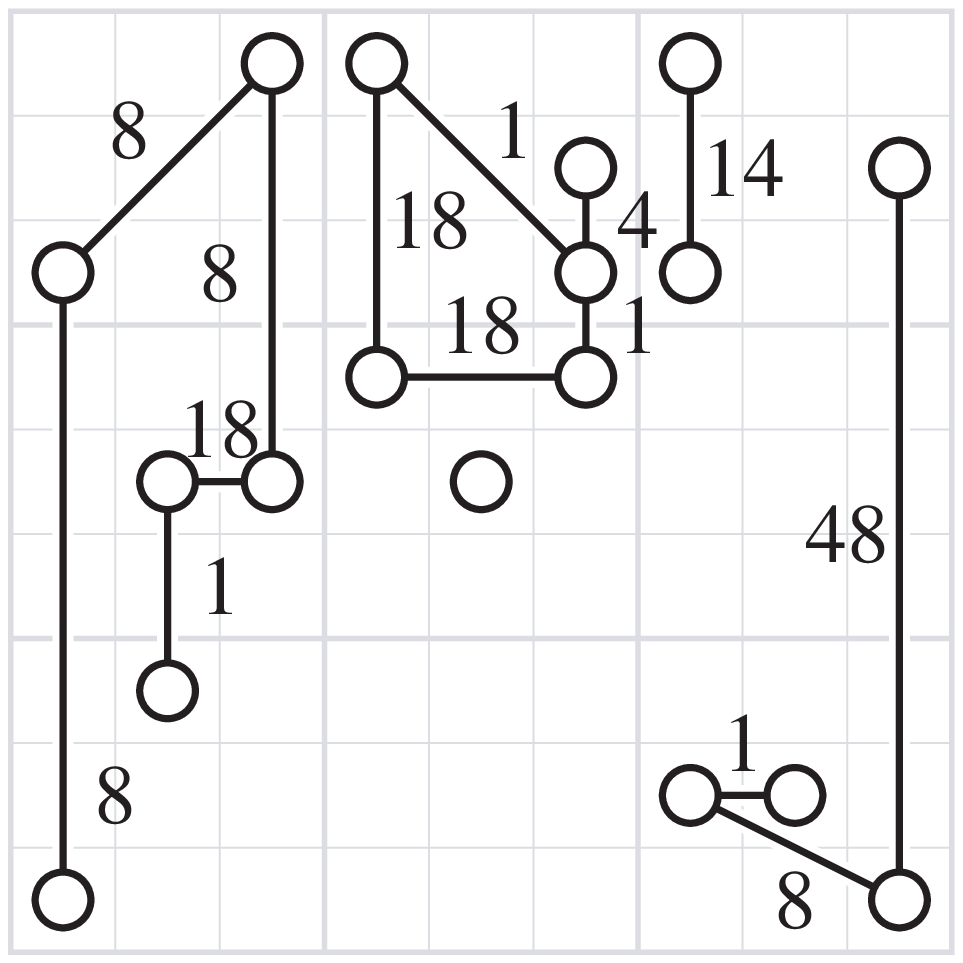}
\end{center}
\caption{A Sudoku grid and its bilocation graph.}
\label{fig:bilocal}
\end{figure}

\begin{figure}[p]
\begin{center}
\includegraphics[width=2.25in]{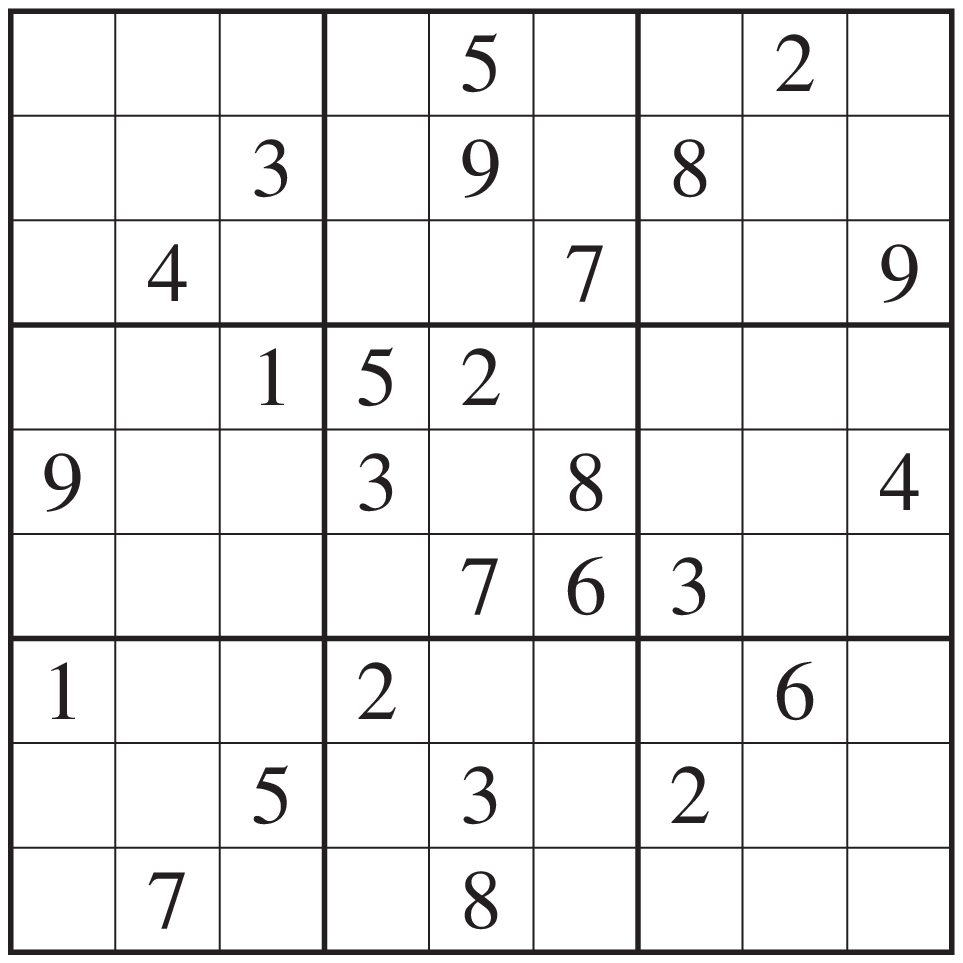}
\hspace{0.25in}
\includegraphics[width=2.25in]{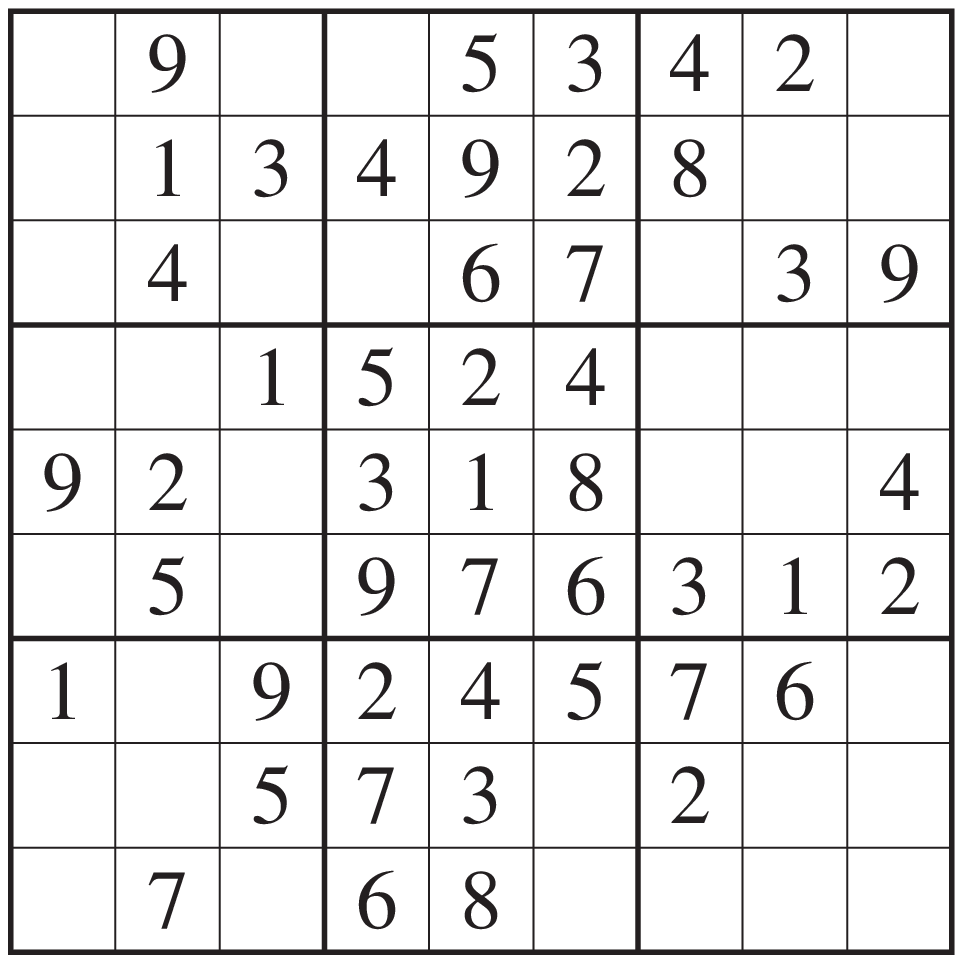}

\vspace{0.25in}

\includegraphics[width=2.25in]{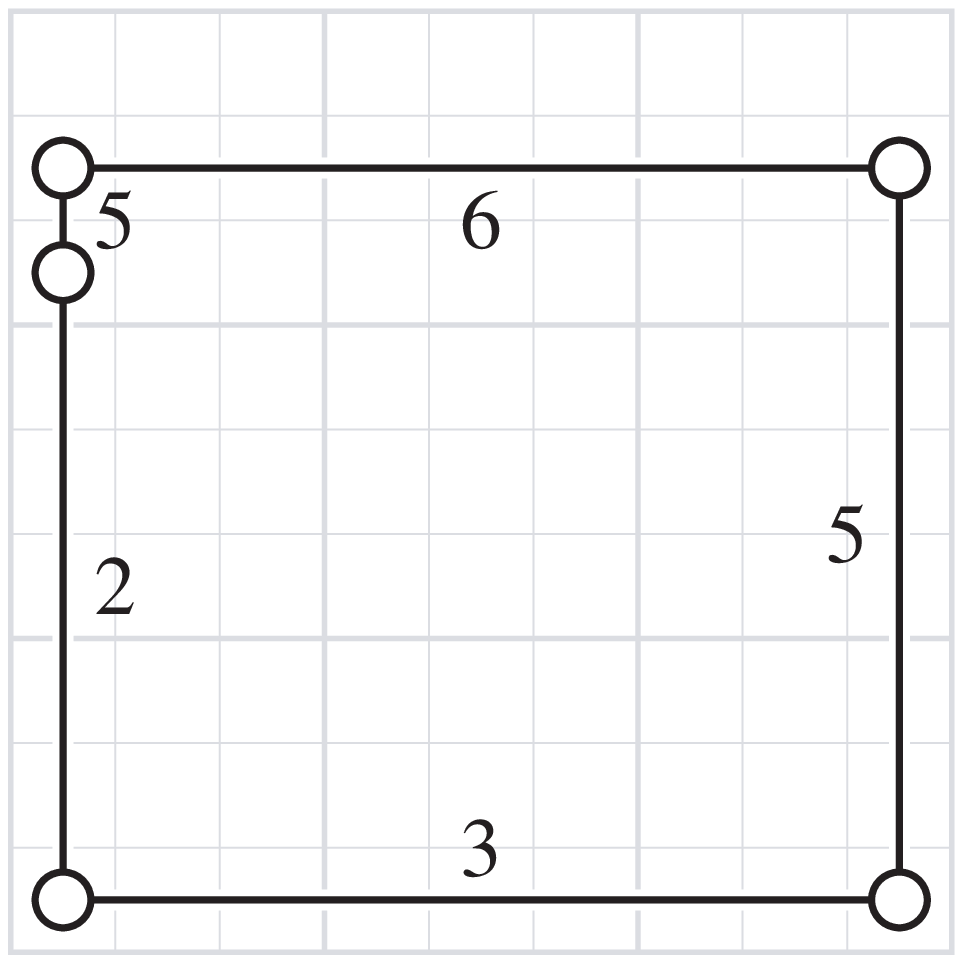}
\hspace{0.25in}
\includegraphics[width=2.25in]{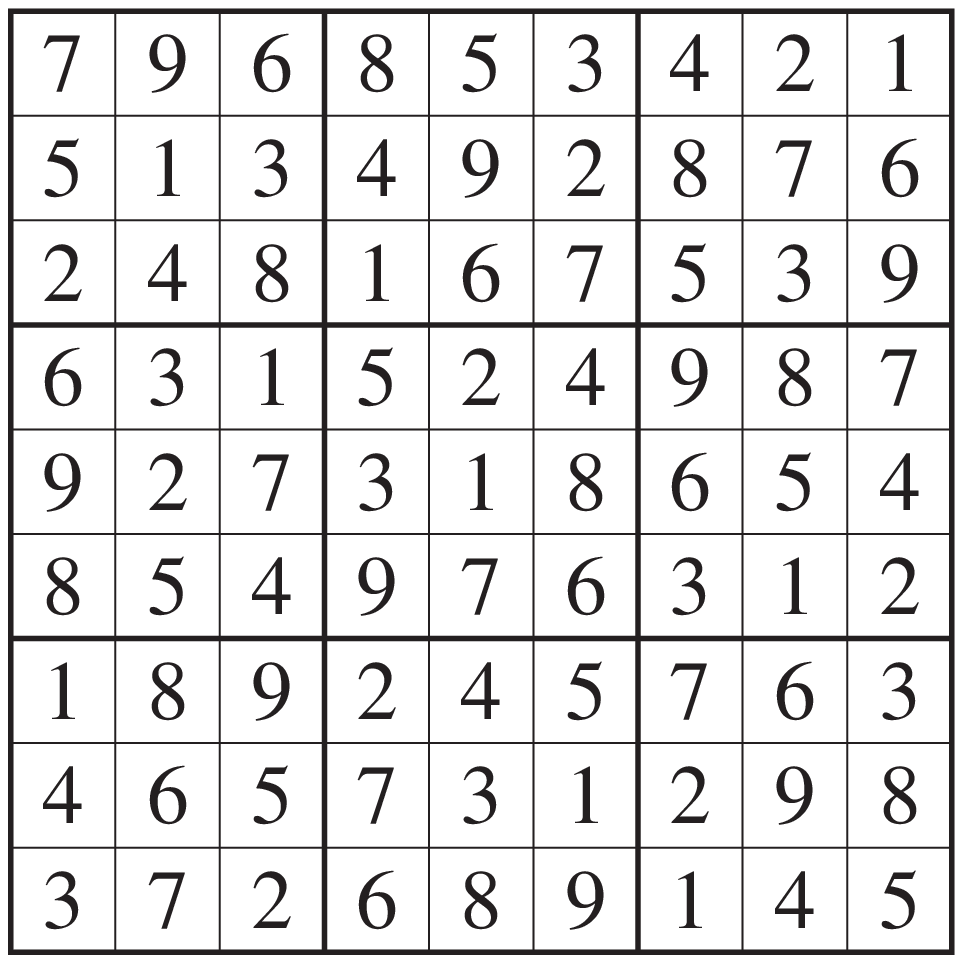}
\end{center}
\caption{Example of the nonrepetitive cycle rule.  The initial state of a Sudoku puzzle is shown on the top left; on top right, we see the same puzzle after the local rules of Section~\ref{sec:local} have been applied to it.  On the bottom left is shown a nonrepetitive cycle in the bilocation graph of the partially solved puzzle.  Due to the existence of this cycle, we can deduce that only the digits 5 or 6 can be placed in the top left and top right empty cells of the cycle, in the second row of the puzzle, leaving only a single cell in that row that can contain the digit 7.  Once the 7 is placed, the remaining puzzle is easy to solve, as shown on the bottom right.}
\label{fig:cycle}
\end{figure} 

In contrast to the local rules described above, our nonrepetitivity-based Sudoku rules are all nonlocal, in that a single application of the rule may consider any cell or digit of the puzzle.  We begin by describing the {\em bilocation graph} of a partially completed puzzle grid, an important tool that helps to visualize the relations among cells and digits of the puzzle.

In the bilocation graph, we draw a vertex for each unfilled cell of the grid.  We connect two vertices by an edge, labeled with a digit $x$, if the corresponding two cells lie in a single row, column, or square of the grid, and if, within that row, column, or square, those two cells have been determined to be the only ones that can contain digit~$x$.  The same pair of cells may be identified as an edge by both a row and a square, or a column and a square; in that case we draw only a single edge.  However, the bilocation graph may have two edges connecting the same pair of vertices, with two different labels.  More than two labels for the same vertex pair would indicate an inconsistency that prevents the Sudoku puzzle from being solved.

An example bilocation graph (for an easy Sudoku configuration) is shown in Figure~\ref{fig:bilocal}.  For instance, the edge in the upper right square of the figure is labeled with the digit~1 because its endpoints are the only ones in their square that can contain that digit, and is also labeled with the digit~4 (so should be thought of as a multiple adjacency) because its endpoints are the only ones in their column that can contain that digit.

\begin{lemma}
The bilocation graph has at most $3B^4$ edges.
\end{lemma}

\begin{proof}
Each of the $3B^2$ rows, columns, and squares of the puzzle can contribute at most $B^2$ edges, one per digit.
\end{proof}

Nonrepetitive paths in the bilocation graph indicate chains of forcing relationships.  For instance, suppose that we have a path of three edges, labeled $x$, $y$, and $z$.  If the first cell of the path is filled in by a digit other than $x$, the first edge forces the second cell to be filled in with an $x$, the second edge in turn forces the third cell to be filled in with a $y$, and the third edge forces the fourth cell to be filled in with a $z$.  However, any single cell of the path may take on a value other than one of the incident edge labels without inconsistency.  Nonrepetitive cycles lead to much stronger restrictions on the cell contents than do nonrepetitive paths:

\begin{lemma}
\label{lem:cycle}
In a partially completed Sudoku puzzle, suppose that cell $c$ belongs to a nonrepetitive cycle in the bilocation graph.  Then $c$ may only be filled with one of the two digits that label its incident edges in the cycle.
\end{lemma}

\begin{proof}
Let those two incident edge labels be $x$ and $y$, and suppose that the solution to the puzzle does not fill $c$ with $x$.  Then by following the forced values around the cycle, as described above, starting with the edge labeled $x$ and continuing around the cycle until $c$ is reached again, we see that $c$ must be filled with $y$.
\end{proof}

Based on Lemma~\ref{lem:cycle}, we define the {\em nonrepetitive cycle rule} as follows.  We identify the set $S$ of all edges that belong to nonrepetitive cycles in the bilocation graph using Theorem~\ref{thm:all-nr}.  For each unfilled cell of the puzzle, we examine the corresponding graph vertex, and collect the set of edge labels belonging to edges of $S$ incident to that vertex.  If this set of labels has three or more digits in it, we have identified an inconsistency that prevents the puzzle from being solved.  If it has two digits, however, we eliminate all other digits as possibilities for the values of that cell.  An example application of the nonrepetitive cycle rule is illustrated in Figure~\ref{fig:cycle}.

If we maintain, as we fill in the puzzle, a set for each digit and each group of cells of the unfilled cells in that group available to that digit, we may identify bilocation edges in total time $O(B^4)$, and the strong connectivity analysis needed to perform the algorithm above also takes this much time.  Therefore, the nonrepetitive cycle rule is a polynomial time rule, as we require for our search procedure.  Indeed, despite its global nature it may be more efficient than several of the local rules we have already defined.

\subsection{Repetitive Cycle Rule}

\begin{figure}[p]
\begin{center}
\includegraphics[width=2.25in]{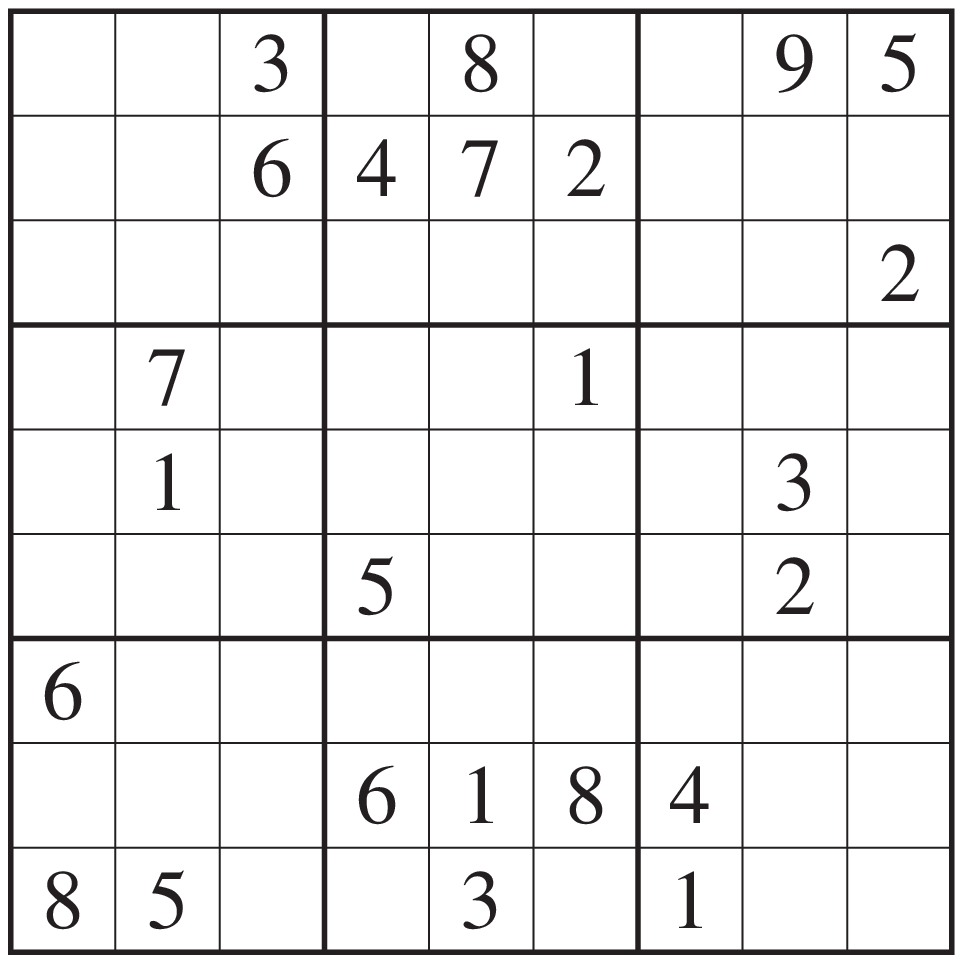}
\hspace{0.25in}
\includegraphics[width=2.25in]{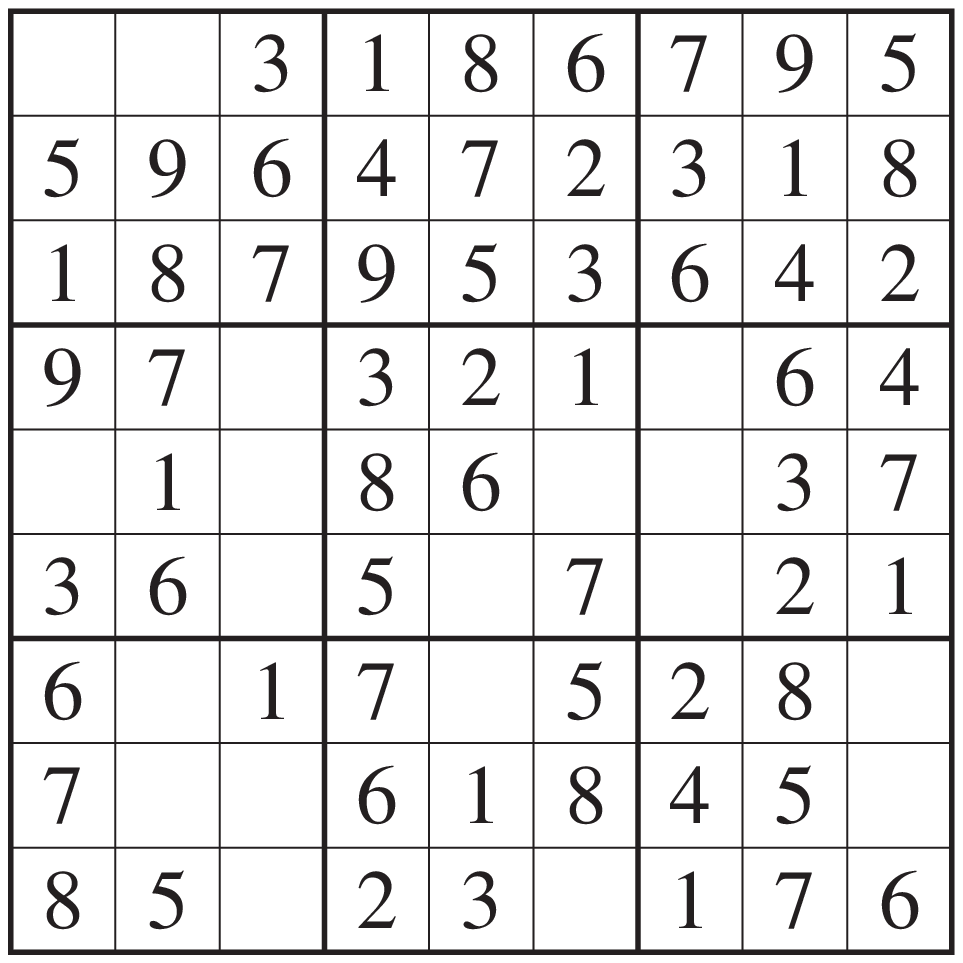}

\vspace{0.25in}

\includegraphics[width=2.25in]{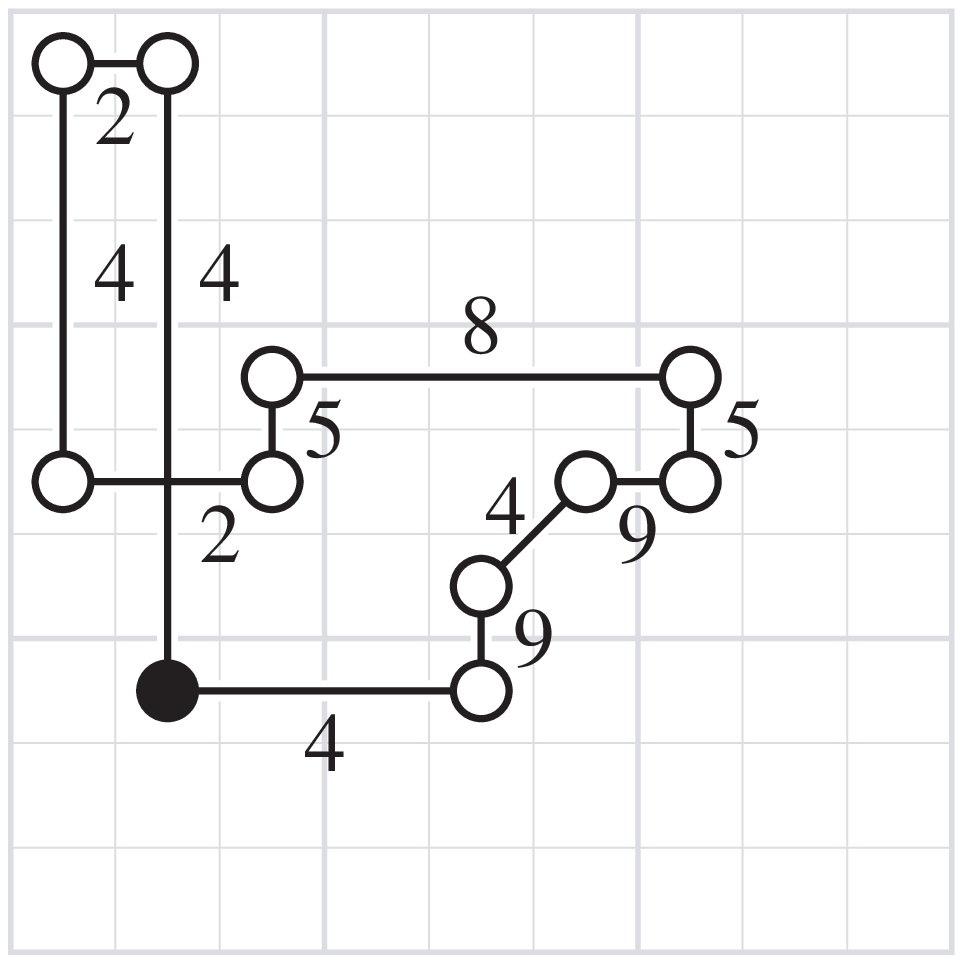}
\hspace{0.25in}
\includegraphics[width=2.25in]{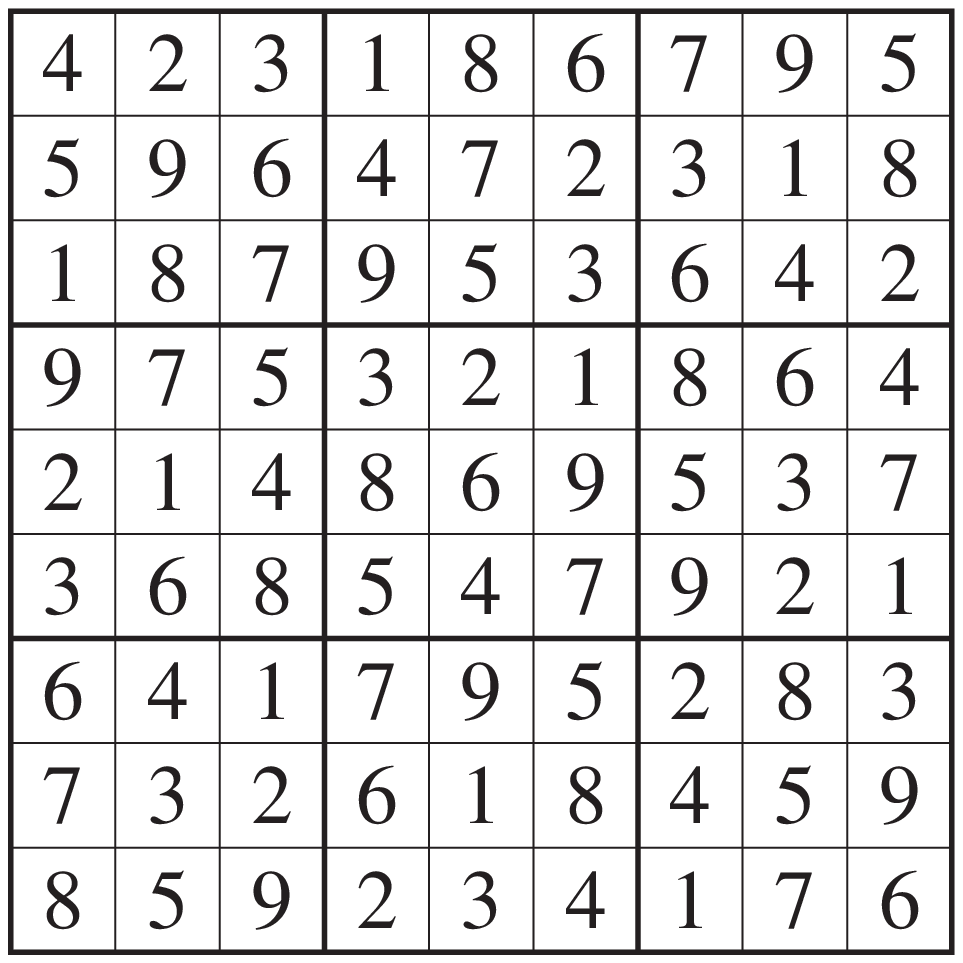}
\end{center}
\caption{Example of the repetitive cycle rule.  The initial state of a Sudoku puzzle is shown on the top left; on top right, we see the same puzzle after the local rules of Section~\ref{sec:local} have been applied to it.  On the bottom left is shown a cycle with one repeated edge label in the bilocation graph of the partially solved puzzle; the repeated label is the 4 on the two edges incident to the solid black vertex.  Due to the existence of this cycle, we can deduce that the cell corresponding to the solid black vertex must contain the digit~4; once this cell is filled, the remaining puzzle is easy to solve, as shown on the bottom right.}
\label{fig:repeat}
\end{figure} 

If the nonrepetitive cycle rule does not make progress on a partially solved Sudoku puzzle, we may still be aided by other cycles in the bilocation graph.

\begin{lemma}
\label{lem:repeat}
Let $C$ be a cycle in the bilocation graph, in which exactly one pair of consecutive edges shares a repeated label.  Then the vertex shared by the edges with the repeated label must correspond to a cell that is filled with that label.
\end{lemma}

\begin{proof}
If the cell were filled with a different digit, its two neighbors in the cycle would have to be filled with the same label, leaving some $k-3$ cells (where $k$ is the length of the cycle) between them to be filled with $k-2$ values.
\end{proof}

Thus, we define the repetitive cycle rule, as illustrated in Figure~\ref{fig:repeat}: search for a cycle $C$ satisfying the criteria of Lemma~\ref{lem:repeat}, and if one is found then fill in the cell indicated by the lemma.  The search can be performed by, for each unfilled cell $c$ and digit $d$, using Theorem~\ref{thm:reach} to test whether there exists a nonrepetitive path that starts and ends with cell $c$ and edge label $d$.  If such a path is found, we fill cell $c$ with digit~$d$.  Each such test takes time $O(B^4)$, so the overall worst case time complexity of this rule is $O(B^8)$.

\subsection{Conflicting Path Rule}

\begin{figure}[p]
\begin{center}
\includegraphics[width=2.25in]{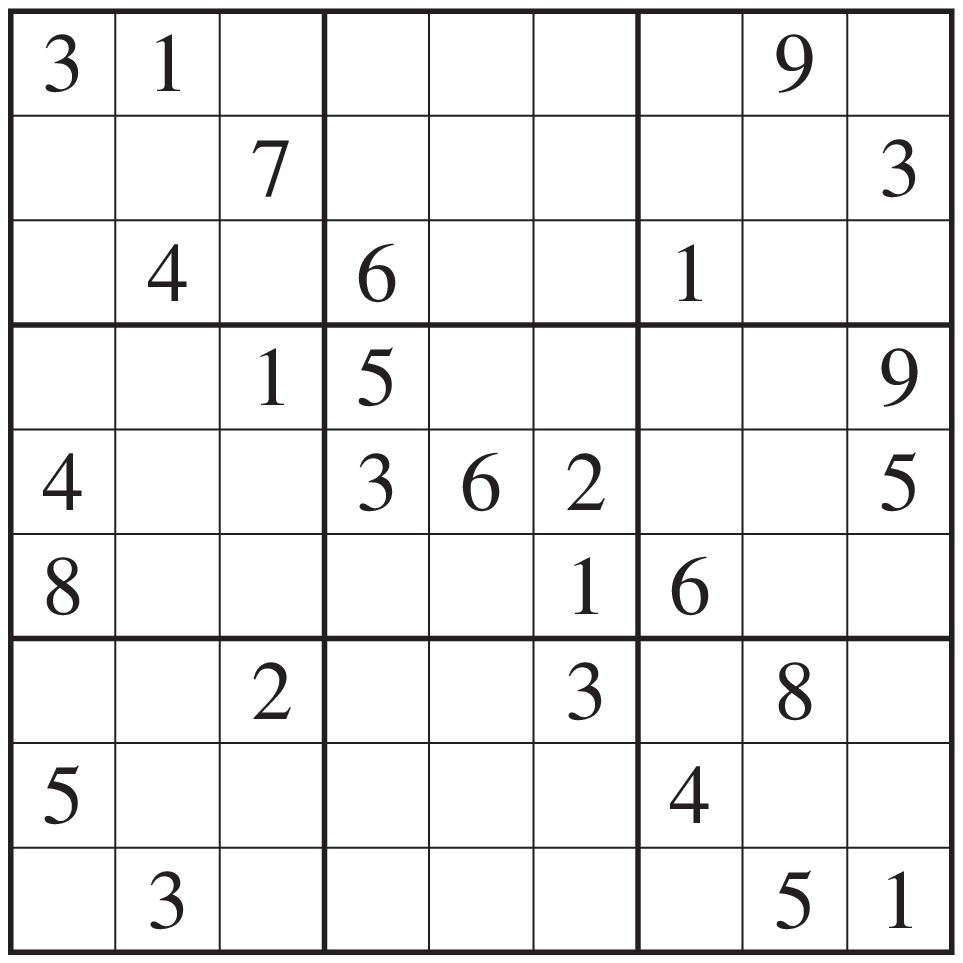}
\hspace{0.25in}
\includegraphics[width=2.25in]{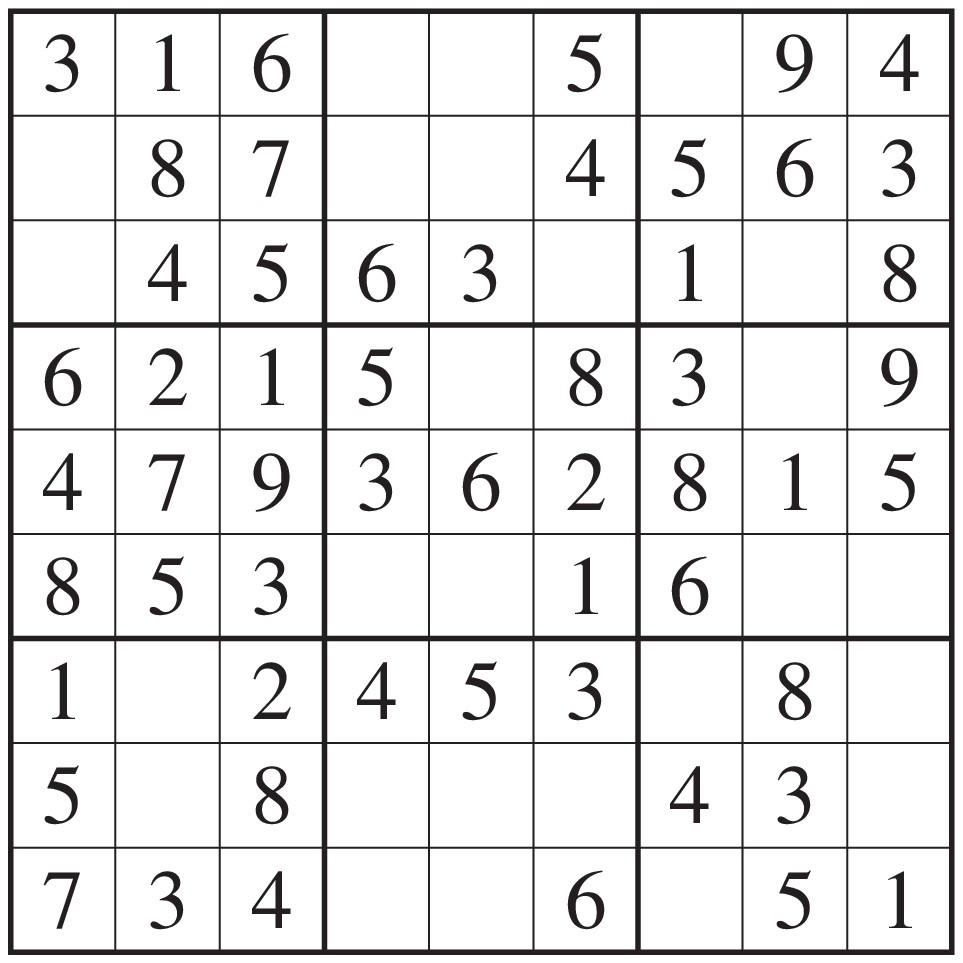}

\vspace{0.25in}

\includegraphics[width=2.25in]{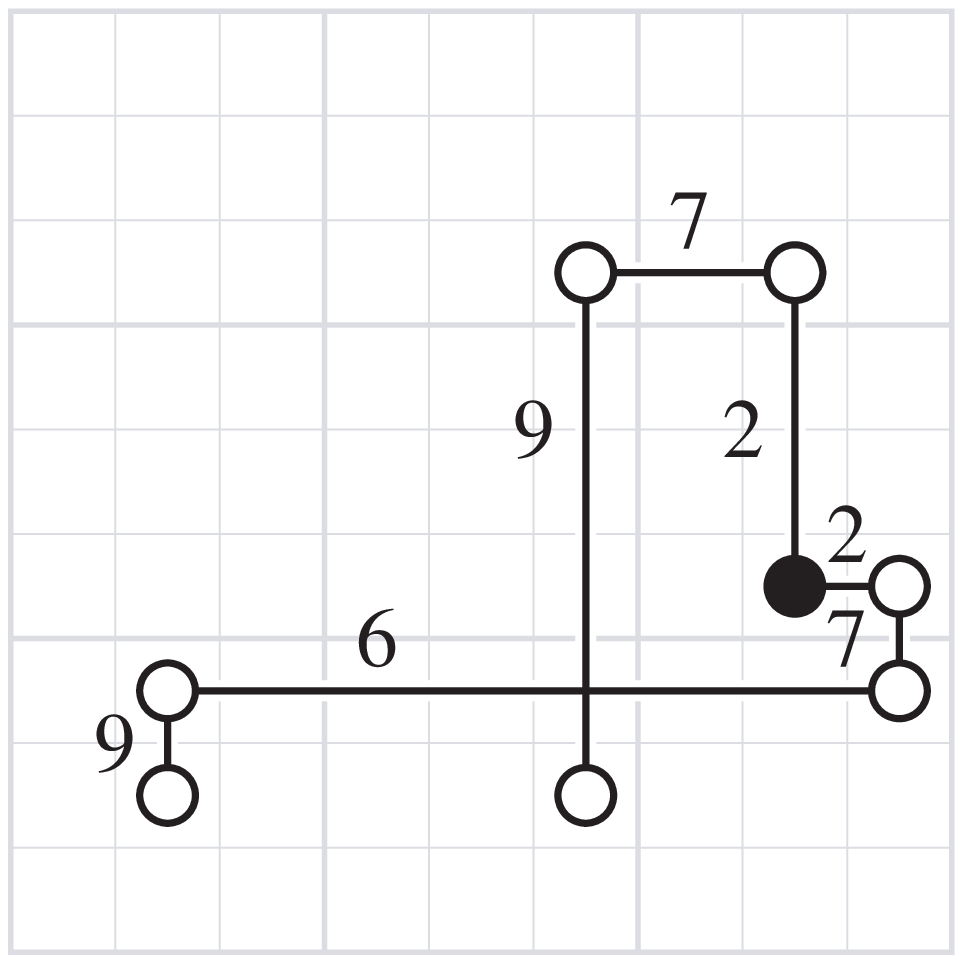}
\hspace{0.25in}
\includegraphics[width=2.25in]{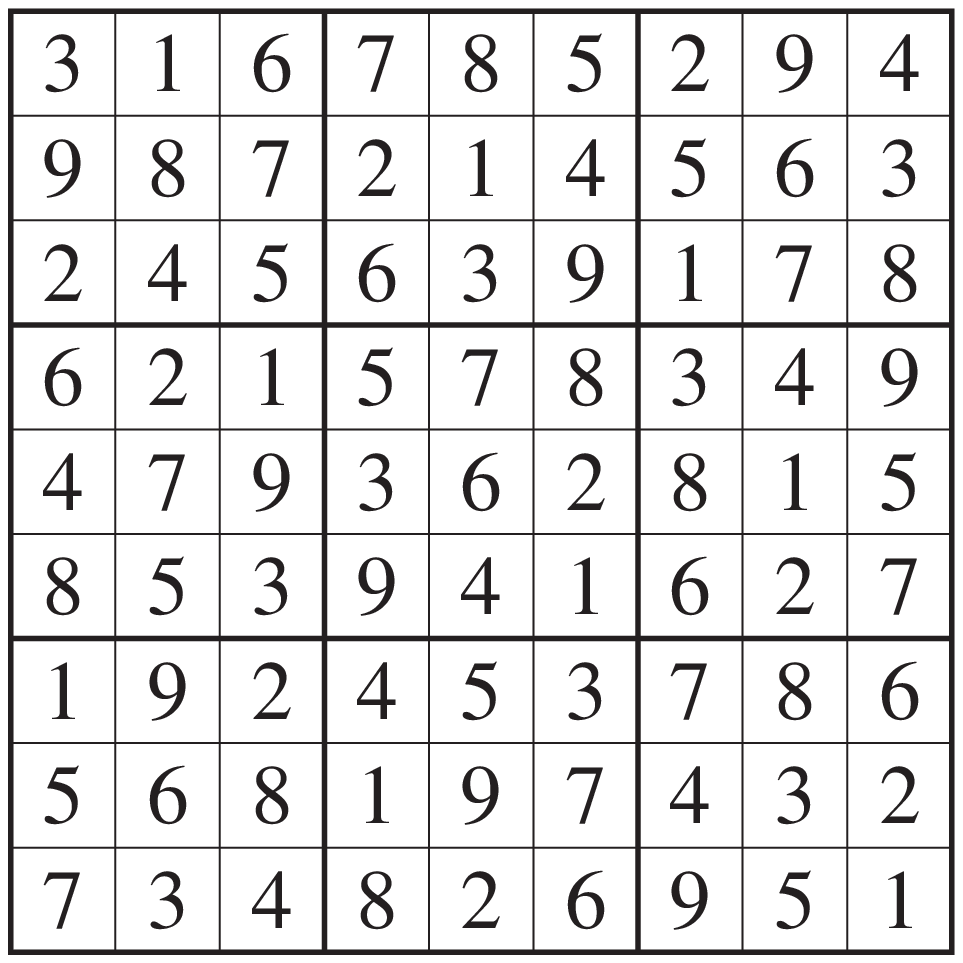}
\end{center}
\caption{Example of the conflicting paths rule.  The initial state of a Sudoku puzzle is shown on the top left; on top right, we see the same puzzle after the local rules of Section~\ref{sec:local} have been applied to it.  On the bottom left are shown two paths in the bilocation graph of the partially solved puzzle, both ending with the same label (9) in the same row.  Due to the existence of these paths, we can deduce that the cell at the start of the paths must contain the start label of the paths,~2.  Once the 2 is placed, the remaining puzzle is easy to solve, as shown on the bottom right.}
\label{fig:conflict}
\end{figure} 

We can generalize the repetitive cycle somewhat, by allowing configurations that resemble repetitive cycles except that one of the links may not even be an edge in the bilocation graph.

\begin{lemma}
\label{lem:conflict}
Suppose $P_1$ and $P_2$ are two paths in the bilocation graph, both beginning with equal labels at the same cell, and both ending with equal labels at distinct cells within the same row, column, or square of the puzzle.  Then, in any solution of the puzzle, the start cell of the paths must be filled with the start label of the paths.
\end{lemma}

\begin{proof}
If this cell were not filled with this label, the forcing relationships represented by the paths would cause the end cells of the paths to have the same label, in conflict with each other.
\end{proof}

The rule based on Lemma~\ref{lem:conflict} is illustrated in Figure~\ref{fig:conflict}.

To search for conflicting pairs of paths, we use Theorem~\ref{thm:reach} to find the pairs of cells and labels reachable by paths from each initial cell and label.  If some label can be the last label on edges to more than $B^2$ cells, some two of these cells must conflict; otherwise we can test for conflicts among the reachable cells for conflicts by bucket sorting them by their rows, columns, and squares, in $O(B^2)$ time.  There are $O(B^4)$ choices for initial cell and label of an incident edge in the bilocation graph, $B^2$ final labels to test, and $O(B^2)$ time per test, so once we have performed the reachability analysis we can do all testing for conflicts in $O(B^8)$ time.  Thus, this is the worst case time bound for the conflicting path rule.

In our actual implementation, we test for conflicts among reachable vertices using bitmasks instead of by sorting; this has a somewhat slower worst case running time but works well in practice.

\subsection{The Bivalue Graph}

\begin{figure}[t]
\begin{center}
\includegraphics[width=2.25in]{graph-example}
\hspace{0.25in}
\includegraphics[width=2.25in]{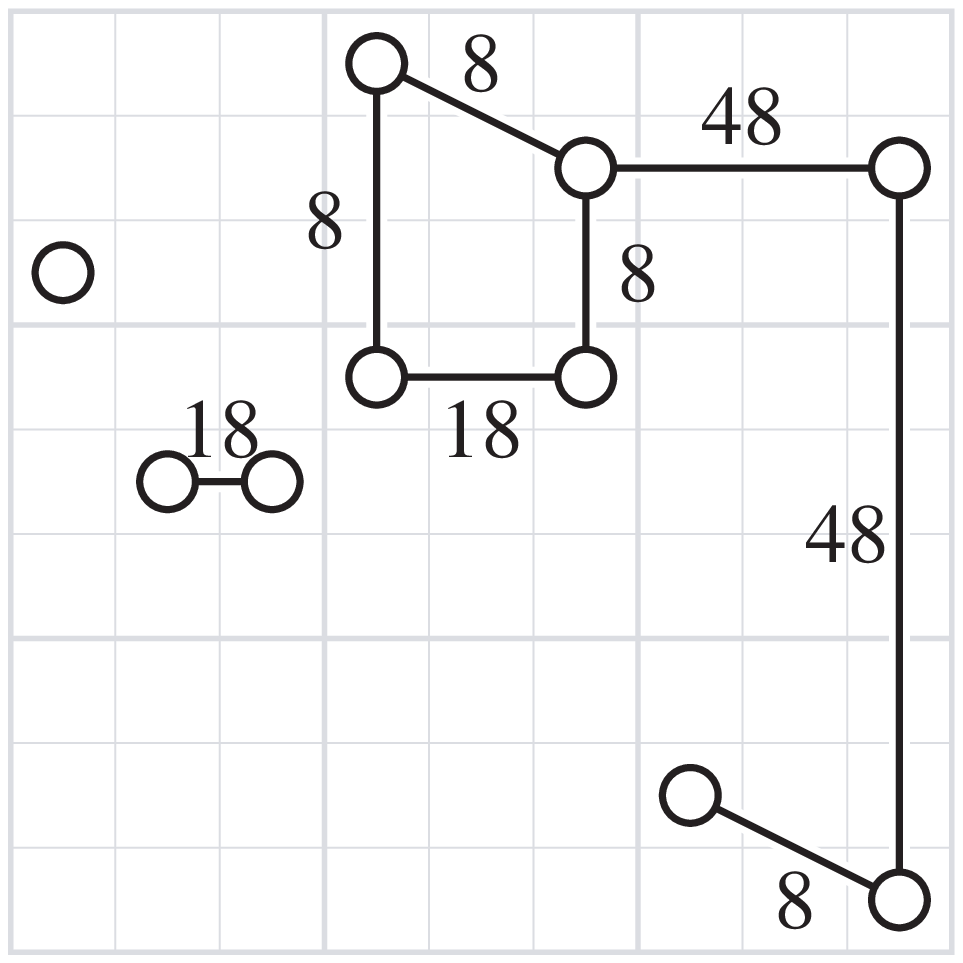}
\end{center}
\caption{A Sudoku grid and its bivalue graph.}
\label{fig:bivalue}
\end{figure}

The rules above have all been based on the bilocation graph of a partially completed Sudoku puzzle.  But there is another graph, the {\em bivalue graph}, that also allows for a similar set of rules.
In this graph, we create a vertex for each cell of the Sudoku grid that has not yet been filled in but for which we have restricted the set of digits that can fill it to exactly two digits.  We connect two such vertices by an edge when the corresponding two cells both lie in a single row, column, or square, and can both be filled by the same digit; the label of the edge is the digit they can both be filled by.
An example bilocation graph (for the same Sudoku configuration that we used as an example for the bilocation graph) is shown in Figure~\ref{fig:bivalue}. 

As with the bilocation graph, nonrepetitive paths in the bivalue graph represent cascading chains of deductions, but in a slightly different way: in the bilocation graph, if the first vertex in the path is not filled with the label of its edge, then the other endpoint is filled with that label, the next vertex in the path is filled with the next label, and so on.  In the bivalue graph, if the first vertex in the path is filled with the label of its edge, then the other endpoint is not, and must be filled with its other possible value, which forms the label of the next edge, and so on.  To describe these cascading deductions in more visual terms, one can imagine that, in the bilocation graph, a cell filled with a mismatched label at the start of the path pushes all the edge labels to the farther cells, while in the bivalue graph, a cell filled with a matched label at the start of the path pulls all the edge labels to the nearer cells.

In any case, the cascading behavior of digit placement in cells of nonrepetitive paths in the bivalue graph lets us define Sudoku rules analogous to those for the bilocation graph:

\begin{itemize}
\item If an edge in the bivalue graph belongs to a nonrepetitive cycle, the digit labeling it must be placed at one of its two endpoints, and can be ruled out as a potential value for any other cell in the row, column, or square containing the edge.
\item If the bivalue graph has a cycle in which a single pair of consecutive edges has a repeated label, that label can not be placed at the cell shared by the two edges, so that cell must be filled by the other of its two possible values.
\item If the bivalue graph contains two paths, both starting with the same label from the same cell,
both ending at cells in the same row, column, or square, and such that in the two ending squares the values not occurring on the incident edge labels are equal, then the cell at the start of the paths can not be filled by the start label of the paths, and must be filled by the other of its two possible values.
\end{itemize}

We can also form a version of the conflicting paths rule that allows one of the two paths to be from the bilocation graph and the other to be from the bivalue graph.

\begin{figure}[t]
\begin{center}
\includegraphics[width=2.25in]{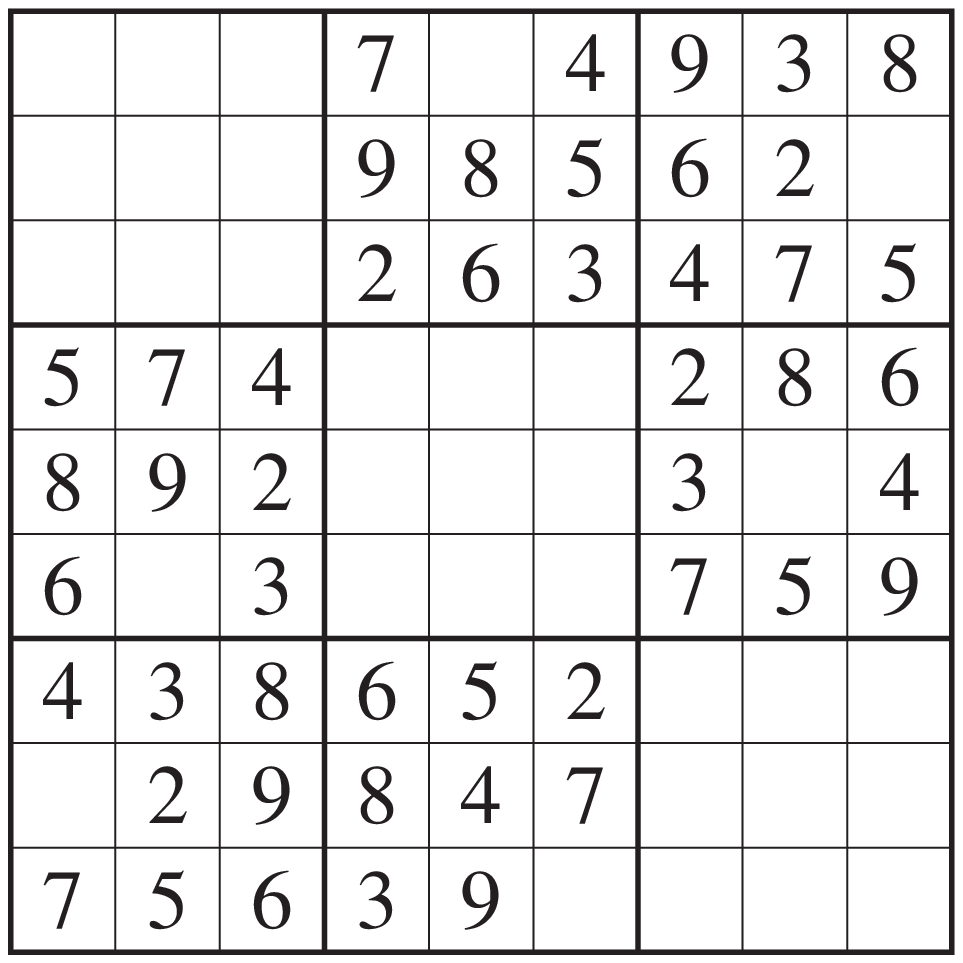}
\end{center}
\caption{$\Omega(B^5)$ lower bound for bivalue graph complexity: if all copies of a single digit (here, the digit~1) and all main diagonal squares are emptied, each of the $B$ emptied squares has $\Omega(B^4)$ bivalue graph edges.}
\label{fig:bival-lb}
\end{figure}

We omit the correctness proofs of these rules, as they are much like those for the bilocation graph.  However, the time analysis is different, as the bivalue graph can be significantly denser than the bilocation graph.  As Figure~\ref{fig:bival-lb} shows, there exist generalized Sudoku puzzle configurations (with a single solution) in which the bivalue graph has $\Omega(B^5)$ edges: start with a filled-in puzzle, empty all cells in the squares along the main diagonal of the puzzle grid, and also remove all copies of a single digit.  Then, in each of the $B$ emptied squares, all but one of the cells is bivalued, with one of the two possible values for each cell being the removed digit, so there are $\Omega(B^4)$ edges per square or $\Omega(B^5)$ overall.  We have not been able to prove any bound on the number of bivalue graph edges better than the obvious $O(B^6)$ that one gets by summing the numbers of edges that can be contained in each row, column, and square of the grid.  It remains an interesting open question how to narrow the gap between $\Omega(B^5)$ and $O(B^6)$ in these bounds, or (of more relevance for actual Sudoku puzzle solution) whether better bounds might be obtained in grids to which no local rules may be successfully applied.

Plugging the bounds on bivalue graph size into the previous analysis for our rules, we see that the bivalue nonrepetitive cycle rule can be implemented to run in $O(B^6)$ time, the bivalue repetitive cycle rule in $O(B^{10})$ time, and the bivalue conflicting paths rule in $O(B^{10})$ time.  In practice these higher time bounds do not seem to be a serious issue with these rules, as the bivalue graph is typically much sparser than our analysis would suggest.  But, from the algorithmic point of view, it is unsatisfactory that these rules have worst case time bounds that are much slower than the similar bilocal graph rules.  Therefore, for the rest of this section, we describe a modification of the bivalue graph and of our repetitivity analysis that allows these rules to be implemented with faster worst-case time bounds.

A {\em flag} in a graph is a pair consisting of a vertex and an edge incident to it; we say that a graph is {\em flag-labeled} if there is a label associated with each flag, and that a path in a flag-labeled graph is nonrepetitive if, for any two consecutive edges, the flags at the vertex they share have different labels.
Our techniques for transforming an edge-labeled graph $G$ into an unlabeled graph $\nr(G)$ such that nonrepetitive paths in $G$ correspond to paths in $\nr(G)$ and vice versa generalize easily to the flag-labeled case.

Define the {\em bipartite bivalue graph} as follows.  There is one vertex per bivalued cell $c$ of the given Sudoku puzzle, and in addition one vertex per pair $(g,d)$ where $g$ is a row, column, or square of the puzzle and $d$ is any digit of the puzzle.  We connect each cell $c$ by six edges, to the pairs $(g,d)$ where row, column, or square $g$ contains cell $c$ and where $d$ is one of the two values that may be placed in cell~$c$.  Thus, the bipartite bivalue graph has at most $4B^4$ vertices and at most $6B^4$ edges.  In addition, for each edge connecting $c$ to $(g,d)$, we label the flag at $c$ by the digit~$d$ and we label the flag at $(g,d)$ by the cell~$c$.  In this way, any nonrepetitive path in the bivalue graph that follows an edge labeled $d$ from $c_1$ to $c_2$ corresponds to a nonrepetitive path in the bipartite bivalue graph that follows two edges, from $c_1$ to $(g,d)$ and from $(g,d)$ to $c_2$, and vice versa.

To implement the bivalue nonrepetitive cycle rule, we can use the flag-labeled version of Theorem~\ref{thm:all-nr} to identify edges that can belong to nonrepetitive cycles of the bipartite bivalue graph, and examine the set of cyclic edges incident to each vertex $(g,d)$.  Every pair of cyclic edges connects $(g,d)$ to two cells belonging to the same strongly connected component of $\nr(G)$, and so can be completed to a cycle; therefore, if there are more than two cyclic edges at $(g,d)$, the application of the cycle rule to each pair would eliminate all cells of $g$ as locations for $d$, showing the puzzle to be inconsistent.  If on the other hand there are exactly two cyclic edges, connecting $(g,d)$ to $c_1$ and $c_2$, we eliminate all cells of $g$ other than $c_1$ and $c_2$ as possible locations for $d$.  In this way, the nonrepetitive cycle rule can be implemented in time $O(B^4)$.  Similarly, by performing nonrepetitive reachability analysis using the flag-labeled version of Theorem~\ref{thm:reach} on the bipartite bivalue graph, we can implement the repetitive cycle rule and the conflicting paths rules in time $O(B^8)$ each.

\subsection{Experimental Results}

We generated a set of 33302 Sudoku puzzles at random using the following procedure:
repeatedly choose at random an unfilled cell, its symmetric partner, and digits to fill those two cells, consistent with previously filled in parts of the puzzle.  Then, use the simplest of our solver rules to propagate the consequences of these choices to other cells of the puzzle.  If we discover an inconsistency, we abort the process and restart it.  Otherwise, the sequence of chosen cells and their values forms an initial state of a Sudoku puzzle that is guaranteed to be solvable by our simplest rules.  Then, in the same order in which we added them, we attempt emptying pairs of cells from this initial state, and use a brute force searcher to test whether the simplified puzzle still has a unique solution.  If it does, we make the removal permanent, and continue testing subsequent pairs.  At the end of this removal process, we have a puzzle that is minimal in the sense that all filled digits are necessary either to preserve a unique solution or to preserve the symmetry of the puzzle's filled cell set.

\begin{figure}[t]
\begin{center}
\includegraphics[width=2.25in]{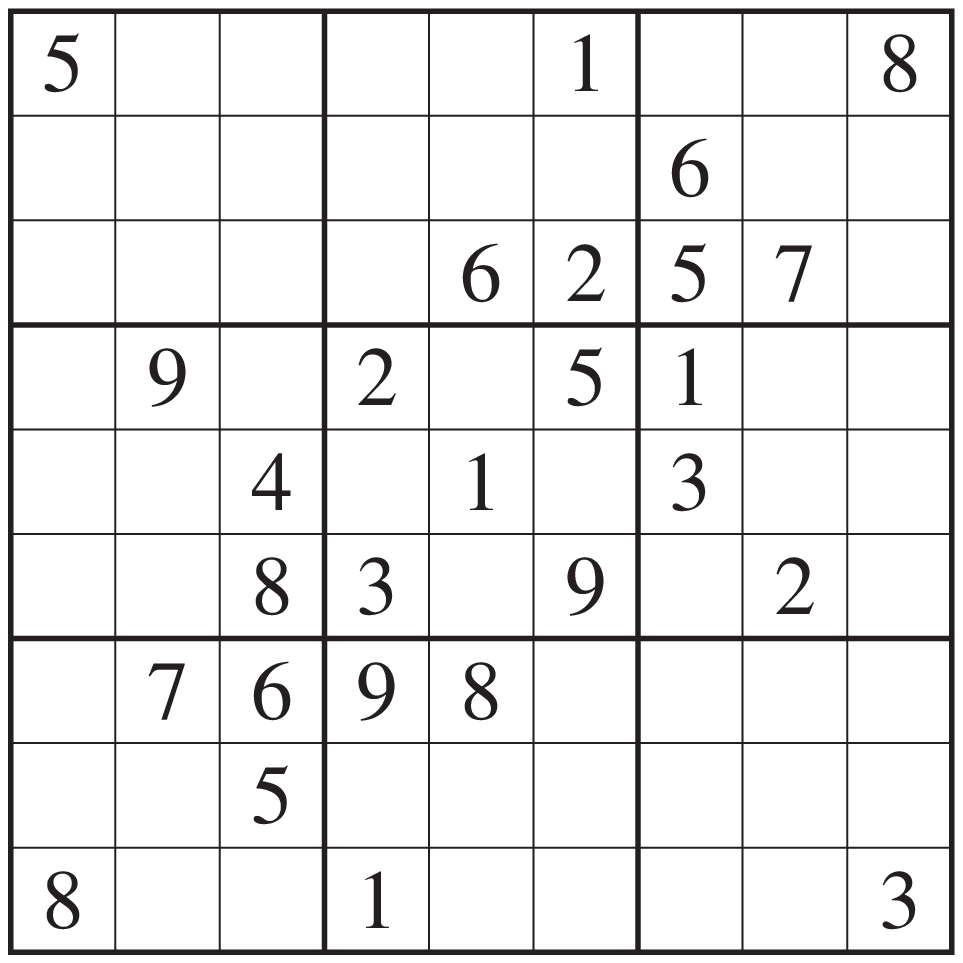}
\end{center}
\caption{A difficult but not unsolvable puzzle generated by our program.}
\label{fig:fiendish}
\end{figure}

We used our non-backtracking solver to test the difficulty of these randomly generated puzzles.  Among them, 1460 puzzles (4.4\%) were unsolvable with our current rule set without backtracking.  Among the remaining solvable puzzles, 3859 of them (11.6\% of the total, or 12.1\% of the solvable puzzles) could be solved only by the use of nonlocal rules based on nonrepetitive paths and cycles, as described in this paper.  Put another way, our nonlocal rules allow us to solve 72.5\% of the problems that would be unsolvable using only local rules.  We conclude that these nonlocal rules significantly reduced the number of unsolvable puzzles generated by our procedure, and can be an effective component in a rule-based Sudoku puzzle solver.  Close examination of the remaining as-yet unsolvable problems may provide inspiration for additional rules that may in future enable our program to solve many more of them.

We leave for the reader to complete the solution to a Sudoku puzzle that is one of the more difficult we generated, but solvable by our rules. The puzzle is depicted in Figure~\ref{fig:fiendish}.

\section*{Acknowledgements}

We would like to thank Jacqueline Hargreaves for helpful discussions in the sudoku livejournal community regarding Sudoku and its automated solution.

\raggedright
\bibliographystyle{abuser}
\bibliography{sudoku}

\end{document}